\documentclass[12pt,fleqn]{article}
\usepackage{amssymb}
\usepackage{amsmath}
\usepackage[dvips]{graphicx,psfrag}
\begin{document}
\begin{flushright}
{\tt hep-ph/0302073 }\\
\end{flushright}
\vspace*{1.5cm}
\begin{center}
    {\baselineskip 25pt
    \Large{\bf 

    Flavor Hierarchy in SO(10) Grand Unified Theories via 5-Dimensional
    Wave-Function Localization
    
    }
    }

\vspace{1.2cm}
\def\thefootnote{\fnsymbol{footnote}}
{\large Ryuichiro Kitano}\footnote
{email: {\tt kitano@ias.edu}}
and 
{\large Tianjun Li}\footnote
{email: {\tt tli@sns.ias.edu}}

\vspace{.5cm}
    
{\small {\it
School of Natural Sciences, Institute for Advanced Study,
Princeton, NJ 08540
}}
    
    \vspace{.5cm}
    \today
    
    \vspace{1.5cm}
    {\bf Abstract}

\end{center}

\bigskip

A mechanism to generate fermion-mass hierarchy in SO(10) Grand Unified
Theories is considered. We find that the lopsided family structure,
which is suitable to the Large angle MSW solution to the solar neutrino
oscillation, is realized without introducing extra matter fields if the
hierarchy originates from the wave-function profile in an extra
dimension. Unlike the Froggatt-Nielsen mechanism, the SO(10) breaking
effect may directly contribute to the source of the hierarchy, 
{\it{i.e.}}, the bulk mass terms. 
It naturally explains the difference of the
hierarchical patterns between the quark and the lepton sectors.
We also find the possibility of the horizontal unification, in which
three generations of the matter fields are unified to a ${\bf 3}$
dimensional representation of an SU(2) gauge group.


\newpage
\def\thefootnote{\arabic{footnote}}
\setcounter{footnote}{0}
\baselineskip 20pt


\section{Introduction}

Supersymmetric Grand Unified Theory (SUSY GUT) is one of the most
promising candidates which describe the interactions among elementary
particles in very high energy phenomena
\cite{Witten:nf}.
SUSY extension of the Standard Model provides us a solution to the
hierarchy problem
\cite{Nilles:1983ge}, 
and the unification of three gauge interactions in
the Standard Model to GUT symmetries explains the charge quantization
which is realized in nature
\cite{Georgi:sy}.  
Moreover, these two scenarios are
non-trivially consistent with each other since SUSY predicts the
coincidence of the gauge coupling constants 
at the high energy scale
\cite{Giunti:ta},
which is a necessary condition to realize the Grand Unification.

Among several candidates for the GUT gauge group such as SU(5), SO(10),
and E$_6$, SO(10) has particularly attractive features.  
One is the matter unification. 
All the quarks and leptons including right-handed
neutrinos in each generation are unified to a single {\bf 16}
dimensional spinor representation field. 
Also, SO(10) is the smallest simple GUT group
which does not require a particular particle content to cancel the
gauge anomaly.
Yukawa unification is another interesting point.  
Because of the matter unification to {\bf 16} representation fields, 
the Yukawa-interaction terms
between the matter (${\bf 16}_i$) and the Higgs (${\bf 10}_H$) fields 
in the superpotential are restricted to the form of 
$W \ni \lambda_{ij} {\bf 16}_i {\bf 16}_j {\bf 10}_H$.
Thus, all the Yukawa matrices for up- and
down-type quarks, the charged leptons, and neutrinos are the same since
they all originate from the above superpotential.
This is not necessarily a desired situation since the above relations
result in the identical masses for the fermions in each family and no
flavor mixings which are completely different from the observed
structure. It is, however, possible to break the above relations
naturally through the non-renormalizable interactions between the matter
fields and ${\bf 45}_H$ Higgs fields which break SO(10) by its Vacuum
Expectation Value (VEV) such as
${\bf 16}_i {\bf 16}_j {\bf 10}_H {\bf 45}_H$,
because the Yukawa interactions may feel the SO(10)
breaking effect
\cite{Albright:1998vf, Babu:1998wi}.

However, when we take into account the fact that the Yukawa matrices
have hierarchical structures and they are quite different among
the quark and lepton sectors, SO(10) GUT has a difficulty in
reproducing such structures.
Many of the recent neutrino oscillation experiments support the Large
Mixing Angle MSW solution 
\cite{Wolfenstein:1977ue} 
(LMA) to the solar neutrino anomaly 
\cite{Fukuda:2001nk}, 
and the large mixing 
in the second and third generation neutrinos is suggested by
atmospheric neutrino oscillation experiments
\cite{Fukuda:1998mi}
and confirmed by the K2K long baseline experiment
\cite{Ahn:2002up}.
On the other hand, all the mixing angles in the quark sector are small
and masses have hierarchical structures. The magnitude of the hierarchy
is enormous, e.g., the ratio of the up-quark mass $m_u$ to the top-quark
mass $m_t$ is approximately $10^{-5}$. Therefore, a mechanism to explain
such a large hierarchy is necessary.
The Froggatt-Nielsen (FN) mechanism is an elegant scenario
to generate hierarchical structures in the Yukawa matrices
\cite{Froggatt:1978nt},
in which the hierarchy comes from the
difference between the fundamental scale $M$ and 
a U(1)$_{\rm FN}$ symmetry
breaking scale $\langle \Phi_{\rm FN} \rangle$.  
The Yukawa matrices $f_u$, $f_d$
and $f_e$, and the neutrino mass matrix $m_\nu$ are given in terms of a
small parameter $\epsilon \sim \langle \Phi_{\rm FN} \rangle / M$ and the
U(1)$_{\rm FN}$
charges $Q$ of the fermions by
$f_u^{ij} \sim \epsilon^{Q(q_i)+Q(u^c_j)}$, 
$f_d^{ij} \sim \epsilon^{Q(q_i)+Q(d^c_j)}$,
$f_e^{ij} \sim \epsilon^{Q(l_i)+Q(e^c_j)}$, and
$m_\nu^{ij} \propto \epsilon^{Q(l_i)+Q(l_j)}$,
where $i$ and $j$ are the generation indices, $q$, $u^c$, $d^c$, $l$,
and $e^c$ represent the corresponding quarks and leptons.
It is known that even if we impose the SU(5) GUT relations among the U(1)
charges, {\it{i.e.}}, $Q(q_i) = Q(u^c_i) = Q(e^c_i) \equiv Q({\bf 10}_i)$ and
$Q(d^c_i) = Q(l_i) \equiv Q({\bf \bar{5}}_i)$, we reproduce the hierarchical
structures for quarks and leptons, and an appropriate neutrino mass
matrix accounting for the solar neutrino anomaly by LMA. 
An example of the
charge assignment is $Q({\bf 10}_1)=3$, $Q({\bf 10}_2)=2$, 
$Q({\bf 10}_3)=0$, $Q({\bf \bar{5}}_3)=1$, $Q({\bf \bar{5}}_2)=0$, 
and $Q({\bf \bar{5}}_1)=0$ with $\epsilon
\sim \lambda \sim 0.22$ 
\cite{Sato:1997hv,Buchmuller:1998zf}.
This type of Yukawa matrices is called lopsided
family structure which gives successful masses and mixing patterns
especially for the large mixing angles in the lepton sector
\cite{Albright:1998vf, Sato:1997hv, Buchmuller:1998zf, Irges:1998ax}.
However, in SO(10) GUT, the above successful mechanism does not work in
a simple way. The hierarchical structures in the fermion masses cannot be
lopsided since the SO(10) symmetry requires 
$Q({\bf 10}_i) = Q({\bf \bar{5}}_i)$.
The situation does not change even if we introduce the non-renormalizable
coupling to ${\bf 45}_H$ to avoid the SO(10) relation in the fermion masses
since the hierarchy is controlled by the charges of 
the U(1)$_{\rm FN}$ symmetry,
which is not related to the SO(10) breaking.
There have been attempts to avoid this problem. An often considered way
is to introduce new matter multiplets of {\bf 10} dimensional
representation and mix or flip $d^c$ and $l$ in the {\bf 16} and those in
the newly introduced {\bf 10} multiplets by the SO(10) breaking effect
\cite{Nomura:1998gm,Maekawa:2001uk,Kitano:2000xk}.
In the context of the six dimensional SO(10) GUT
\cite{Asaka:2001eh}, 
it has been also
proposed the introduction of extra matter fields where the hierarchy is a
consequence of the volume suppression factors
\cite{Haba:2002ve}.
In these scenario, $d^c$ and $l$ originate from the {\bf 10}
representation fields other than {\bf 16}, 
and thus the matter
unification is spoiled.
This is a generic problem in SO(10) GUT. 
Therefore, in order to solve this problem,
the strong connection between the origin of flavor hierarchy and the
SO(10) breaking is necessary.

In this sense, the recently proposed mechanism to generate the flavor
hierarchy based on 
the wave-function profiles in
the extra dimension is noteworthy
\cite{Arkani-Hamed:1999dc, 
Mirabelli:1999ks, Kaplan:2001ga, Kakizaki:2001ue, Haba:2002uw}.  
In the simplest scenario, 
our world has an $S^1/Z_2$ compactified extra spacial dimension,
and the bulk mass terms for the bulk superfields make them localize on
the branes at the orbifold fixed points
\cite{Kaplan:2001ga,Arkani-Hamed:2001tb}. 
Consequently,
the values of the
wave functions at a brane may be suppressed by a factor of $\exp(-|m| \pi
R)$, where $m$ is the bulk mass and $R$ is the radius of the extra
dimension. The flavor hierarchy is obtained in a similar way as the
FN scenario such like $f^{ij} \sim \exp(-(|m_i| + |m_j|) \pi
R)$.
Hebecker and March-Russell considered the scenario in SU(5) GUT and
showed that the hierarchy can be reproduced with a natural parameter
sets
\cite{Hebecker:2002re}.
Again, in SO(10), it seems that the SO(10) relation, {\it{i.e.}}, the same
$m_i$ for all the matter fields in each family, is too strong to
reproduce the hierarchy.  
However, it is possible to break the SO(10) relations
since we have
additional contributions to the bulk masses $m$'s when SO(10) is broken
by the VEV of the bulk adjoint field whose existence is ensured by SUSY
in five dimension. The additional contributions are not universal to all
the matter fields but proportional to the U(1)$_X$ ($\subset$
SU(5)$\times$U(1)$_X$ $\subset$ SO(10)) charges, so that they can change
the wave-function profiles without preserving the SO(10) relations.

In this paper, we construct an SO(10) SUSY GUT model in 5-dimensional
space-time in which the correct fermion-mass patterns are realized by
the above mentioned mechanism. The matter fields in each family are
unified to a single {\bf 16} representation field. Our scenario is
compatible with already proposed doublet-triplet splitting mechanisms
such as Dimopoulos-Wilczek mechanism
\cite{Dimopoulos-Wilczek}, 
boundary condition
\cite{Kawamura:2000ev,Hebecker:2001wq}, 
etc. 
We also discuss a possibility of the horizontal unification 
which unifies the three families 
to the single three dimensional representation
of SU(2) group.

This paper is organized as follows: In Section 2, we construct a model
and explain the mechanism. The parameter sets to realize the suitable
fermion-mass patterns are discussed in Section 3. Section 4 is devoted to
discussions and conclusions.

\section{Model}

In this section, we first review SUSY theory on the space-time $M^4
\times S^1/Z_2$, and show that the 5-dimensional wave function of zero
mode has an exponential shape.
This mechanism is a key ingredient of the fermion mass hierarchy.
Then, we discuss the SO(10) models in this setup.

\subsection{Zero mode wave function due to bulk mass term}

We consider the $S^1/Z_2$ compactified 5-dimensional SUSY gauge
theory 
\cite{Pomarol:1998sd}. 
For our purpose, it is convenient to write
down the action in terms of superfields in 4-dimensional superspace
\cite{Arkani-Hamed:2001tb}. 
The action for the U(1) gauge theory with a
bulk hypermultiplet is given by
\begin{eqnarray}
 S \!\!\!&=&\!\!\! \int d^4 x \int_0^{\pi R} dy
\left[ {\rule[-3mm]{0mm}{10mm}\ } \right.
\int d^2 \theta \left(
\frac{1}{4} W^\alpha W_\alpha + {\rm h.c.}
\right)
+ \int d^4 \theta \left(
\partial_y V - \frac{1}{\sqrt{2}} ( \phi + \bar{\phi} )
\right)^2 
\nonumber \\
&+&\!\!\!  \int d^4 \theta
\left( H^c e^{2 g Q V} \bar{H}^c 
+ \bar{H} e^{-2 g Q V} H \right)
+ \int d^2 \theta
\left(
H^c ( m + \partial_y - \sqrt{2} g Q \phi ) H
 + {\rm h.c.}
\right)
\left. {\rule[-3mm]{0mm}{10mm}\ }\right]
\ ,
\nonumber \\
\label{action}
\end{eqnarray}
where $W^\alpha$ and $V$ are the field strength and vector superfield
associated with the U(1) gauge group, and $\phi$ is the chiral
superfield where $A_5$ is in its lowest components.
The gauge transformation for the superfield is
$\phi \to \phi + 1/({\sqrt{2}g}) \partial_y \Lambda$ 
with transformation parameter $\Lambda$. 
Note that the gauge coupling constant $g$ has a mass
dimension of $-1/2$.  The chiral superfields $H$ and $H^c$ are the
components of the hypermultiplet and their U(1) charges are $Q$ and
$-Q$, respectively\footnote{The superfields $H$ and $H^c$ do not mean
the Higgs fields here.}.  
To be invariant under the orbifold projection, the
$Z_2$ parity is assigned to be even for $W^\alpha$, $V$, and $H$, and
odd for $\phi$ and $H^c$. 
Also, the bulk mass parameter $m$ should be $Z_2$
odd such that $m = m \ {\rm {sgn}}(y)$.

By the Kaluza-Klein (KK) decomposition, we can find the existence of a
zero mode with a localized wave function.
We expand the chiral superfields $H$ and $H^c$ as follows:
\begin{eqnarray}
 H(y) = \sum_{n=0}^{\infty} H_n(x) f_n(y) \ ,\ \ \ 
 H^c(y) = \sum_{n=0}^{\infty} H^c_n(x) g_n(y)\ .
\end{eqnarray}
To obtain the normalized kinetic terms in the 4-dimensional effective
action, the functions $f_n(y)$ and $g_n(y)$ are solutions of the
following differential equations:
\begin{eqnarray}
 (\frac{d}{dy} + m) f_n(y) = m_n g_n(y) \ ,\ \ \ 
 (- \frac{d}{dy} + m) g_n(y) = m_n f_n(y)\ ,
\end{eqnarray}
with the normalization conditions:
\begin{eqnarray}
 \int_0^{\pi R} dy \ f_n(y) f_m(y) =
 \int_0^{\pi R} dy \ g_n(y) g_m(y) = \delta_{nm}\ .
\end{eqnarray}
The zero mode wave function with correct $Z_2$ parity
is easily found to be: 
\begin{eqnarray}
 f_0(y) = \sqrt{\frac{2m}{1-e^{-2m \pi R}}} \ e^{-my}\ ,\ \ \ 
 g_0(y) = 0\ .
\label{zero-mode}
\end{eqnarray}
Thus the zero mode wave function $f_0$ localizes exponentially at $y=0$
$(\pi R)$ for $m>0$ ($m<0$).  For $m<0$ and $-2m \pi R \gg 1$, since the
value of the wave function at $y=0$ is exponentially suppressed by a
factor of $f_0(0) \sim \sqrt{2|m|} \ e^{- |m| \pi R}$, the strength of
the couplings between the bulk fields and the brane fields on the $y=0$
brane becomes weak.
This effect can be the origin of the fermion mass hierarchy if the Higgs
fields are confined on the brane at $y=0$
\cite{Kaplan:2001ga}.

For the massive modes, the KK masses are given by
\begin{eqnarray}
 m_n^2 = m^2 + \left( \frac{n}{R} \right)^2 \ .
\end{eqnarray}

\subsection{Gauge symmetry breaking}

As we can see in Eq.(\ref{action}), the VEV of the $\phi$ superfield is
an additional contribution to the bulk mass $m$, 
and as a consequence the
wave-function profile of the zero mode is changed
\cite{Kaplan:2001ga,Arkani-Hamed:2001tb,Barbieri:2002ic}. 
Note that the additional contribution is proportional to the U(1) charge.
We identify the U(1)
gauge symmetry as a subgroup of the SO(10) GUT gauge group and $H$ as a
matter superfield of the {\bf 16} dimensional representation of SO(10),
and then the different wave-function profiles are realized for the matter
fields embedded in {\bf 16} since all the components do not have the
same U(1) charges.

We consider the vacuum where the $\phi$ superfield acquires the VEV
while the matter superfields $H$ and $H^c$ do not.
In order to preserve $N=1$ SUSY in the 4-dimensional effective theory
(it ensures the configuration to be true vacuum), the VEVs of $F$- and
$D$-terms must vanish. The vanishing $F$-term conditions are satisfied by
assuming $\langle H \rangle = \langle H^c \rangle = 0$, the $D=0$
condition is not the same as the 4-dimensional case but as follows
\cite{Kaplan:2001ga,Arkani-Hamed:2001tb,Barbieri:2002ic}:
\begin{eqnarray}
 0 = -D = \partial_y \langle \phi \rangle \ ,
\end{eqnarray}
where we denote that $\phi$ is its scalar component. 
The contribution from the $\phi$ field is due to the 5-dimensional
kinetic term, {\it{i.e.}}, the second term in Eq.(\ref{action}).
The above equation is automatically satisfied if $\phi = {\rm const.}$,
however, as mentioned earlier, the $\phi$ field has odd parity under $y
\to -y$ transformation which means that the VEV should have the form of
$\langle \phi \rangle = v^{3/2} \ {\rm sgn}(y)$. Substituting this form
in the above equation, we obtain the non-vanishing $D$-terms on the
branes as follows:
\begin{eqnarray}
 -D = 2v^{3/2} \left( \delta(y) - \delta(y-\pi R) \right) \ .
\label{D-term}
\end{eqnarray}

The brane localized $D$-terms in Eq.(\ref{D-term}) have to be canceled
by the brane localized terms in the action.
The most economical way is to add the Fayet-Iliopoulos $D$-terms on 
both branes. However, it is not applicable in our context since we
identify the U(1) symmetry as a subgroup of SO(10)\footnote{ The
$D$-term cancellation by the Fayet-Iliopoulos term is possible if the
gauge group is broken explicitly to SU(5)$\times$U(1) on the brane by
the boundary conditions.}.
The second way is to add the brane fields charged under the U(1)
symmetry. For example, if we assume the presence of chiral superfields
$S_0$ with charge $q$ and $S_\pi$ with charge $-q$ which are localized
on the branes at $y=0$ and $\pi R$, respectively, the $D$-term is
\begin{eqnarray}
 -D = \partial_y \langle \phi \rangle 
-gq |\langle S_0 \rangle|^2 \delta(y)
+gq |\langle S_\pi \rangle|^2 \delta(y-\pi R)\ .
\end{eqnarray} 
The condition $D=0$ has a solution with a flat direction:
\begin{eqnarray}
 2 v^{3/2} = gq|\langle S_0 \rangle|^2 
= gq|\langle S_\pi \rangle|^2 \ .
\end{eqnarray}
This is the usual $D$-flat direction in the 4-dimensional
effective theory.
In the vacuum with $v \neq 0$,
the U(1) symmetry is
broken by the VEVs of $S_0$ and $S_\pi$. 
This is an important point when we apply this gauge symmetry breaking
to the SO(10) GUT breaking.  The fact that the U(1) symmetry is broken
indicates that the U(1) symmetry should not be included in the Standard
Model gauge group. Moreover, the U(1) symmetry must be orthogonal to the
Standard Model gauge group,
otherwise the VEV of $\phi$ induces the
$D$-term corresponding to the Standard Model gauge group which cannot be
canceled without their breaking.
Therefore, the U(1) subgroup of SO(10) is uniquely determined to be so
called U(1)$_X$ symmetry, and thus SO(10) is broken down to SU(5). As we
see later, this U(1)$_X$ breaking yields the correct masses and mixing
patterns of the fermions through the wave-function deformation as well
as the Majorana masses for the right-handed neutrinos.

\subsection{SO(10) model}

We construct an SO(10) GUT model which includes the above setup. The
matter fields are unified to three copies of {\bf 16} representation,
${\bf 16}_1$, ${\bf 16}_2$, and ${\bf 16}_3$ of SO(10), and they and
their charge conjugation fields ${\bf 16}^c_i$ can propagate into the
bulk. Also, the gauge fields including a chiral superfield ${\bf
45}_\phi$ as the $\phi$ component in the action Eq.(\ref{action}) are in
the bulk. The Higgs field ${\bf 10}_H$ is introduced as a brane field
(on the $y=0$ brane) in order to obtain the suppressed Yukawa coupling
constants for the first and second generations.
To break SO(10) GUT, we need ${\bf 45}_H$ Higgs field
which has VEV along the $B-L$ direction 
in the Dimopoulos-Wilczek mechanism
\cite{Dimopoulos-Wilczek},
and the Higgs
fields of ${\bf 16}_H$ and ${\bf \overline{16}}_H$ are necessary to
reduce the rank of gauge group from five to four and give the
Majorana masses to the right-handed neutrinos. We put these fields on
the $y=0$ brane, and ${\bf 16}_H^\prime$ and 
${\bf \overline{16}}_H^\prime$ on the $y=\pi R$ brane so that they can play
the same roles as $S_0$ and $S_\pi$ in the above mechanism.

In this setup, the superpotential is given by
\begin{eqnarray}
 W &=& {\bf 16}^c_i ( m_i + \partial_y - \sqrt{2} g {\bf 45}_\phi ) 
{\bf 16}_i
\nonumber \\
&+&
\frac{\delta (y)}{M} \left[
\lambda_{ij} {\bf 16}_i {\bf 16}_j {\bf 10}_H
+ 
\frac{\tilde{\lambda}_{ij}}{M}
({\bf 16}_i {\bf \overline{16}}_H)_{\bf 1} 
({\bf 16}_j {\bf \overline{16}}_H)_{\bf 1}
\right]
\nonumber \\
&+& \delta (y)
V_0({\bf 45}_H, {\bf 16}_H, {\bf \overline{16}}_H, {\bf 10}_H, \cdots)
\nonumber \\
&+&
\delta (y-\pi R) 
V_\pi({\bf 16}_H^\prime, {\bf \overline{16}}_H^\prime)
\ ,
\label{superpotential}
\end{eqnarray}
where $i=1,2,3$ are the generation indices and $M$ is a parameter with
mass dimension one. 
The first term is the required form from the SUSY in five dimension.
We take a flavor basis where the bulk mass terms $m_i$ are diagonalized.
The Yukawa coupling constants $\lambda_{ij}$ are dimensionless
quantities and the interaction terms with $\tilde{\lambda}_{ij}$ induce
the Majorana masses for the right-handed neutrinos after the SO(10)
breaking. We omit the other possible contractions which are irrelevant
for the low-energy effective theory. 

The potential $V_0$ needs to have an appropriate form for the
Dimopoulos-Wilczek mechanism. 
According to the simplest model in Ref.\  
\cite{Barr:1997hq}, 
it is required to exist another ${\bf 10}_{2H}$, an additional pair of
${\bf 16}_{2H}$ and ${\bf \overline{16}}_{2H}$, and
several singlet fields\footnote{It is difficult to forbid the dimension
five operators which cause the proton decay in SO(10) model with
the Dimopoulos-Wilczek mechanism. The suppression of these
operators is discussed in Ref.\
\cite{Babu:1998wi,Maekawa:2001uk,Babu:1993we}.}.
On the $y=\pi R$ brane, we can write the arbitrary potential $V_\pi$ for the
${\bf 16}_H^\prime$ and ${\bf \overline{16}}_H^\prime$, and in general,
there are a number of vacua where SO(10) is broken.
The Standard Model singlet components $N_H$ and $\bar{N}_H$ ($\in {\bf
16}_H$ and ${\bf \overline{16}}_H$),
and 
$N_H^\prime$ and $\bar{N}_H^\prime$ 
($\in {\bf 16}_H^\prime$ and ${\bf \overline{16}}_H^\prime$) 
acquire VEVs after minimizing the
potential $V_0$ and $V_\pi$.
Since these singlets are charged under
U(1)$_X$ subgroup of SO(10), there is a non-trivial condition from the
$D$-flatness which is given by
\begin{eqnarray}
 0 = -D_{\rm U(1)_X} = \!\!\!&&\!\!\! \delta (y) \left[
v^{3/2} + 5 g_X (|\langle N_H \rangle |^2 
- |\langle \bar{N}_H \rangle |^2)
\right] \nonumber \\
\!\!\!&&\!\!\!
- \delta (y-\pi R)
\left[
v^{3/2} - 5 g_X (|
\langle N_H^\prime \rangle |^2 - |
\langle \bar{N}_H^\prime \rangle |^2)
\right]\ ,
\label{d-flat}
\end{eqnarray}
where $g_X$ is the coupling constant of the U(1)$_X$ gauge
interaction normalized such that the charge of $N_H$ is $-5$, and
$v^{3/2}$ is the VEV of the ${\bf 45}_\phi$ field in the U(1)$_X$
direction.
Along the flat direction in Eq.(\ref{d-flat}), we obtain the vacuum
where $v \neq 0$ as seen in the previous subsection.

Even with the potential $V_\pi$, there are massless modes because of the
global SO(10) symmetry in the $y=\pi R$ sector. We assume the modes are
stabilized by the SUSY breaking effect.
The minimal set of the massless modes are a pair of ${\bf 10}$ and ${\bf
\overline{10}}$ of SU(5) and a Standard Model singlet since the VEVs of
$N_H^\prime$ and $\bar{N}_H^\prime$ break SO(10) to SU(5), although it
requires additional Higgs fields. The massless modes will obtain the
masses of the order of TeV.
Alternatively, the singlet mode can be stabilized without SUSY breaking
effect.
When we gauge the U(1)$^\prime$ symmetry under which only ${\bf
16}_H^\prime$ and ${\bf \overline{16}}_H^\prime$ are charged such as
${\bf 16}_H^\prime:1$ and ${\bf \overline{16}}_H^\prime:-1$,
the constraint from the
$D$-flat condition is given by
\begin{eqnarray}
 0 &=& -D_{\rm U(1)^\prime} 
\nonumber \\
&=& \delta (y) \left[
\xi_0 + v^{\prime 3/2} 
\right]
-\delta (y - \pi R) \left[
- \xi_\pi + v^{\prime 3/2} 
- g^\prime (|
\langle N_H^\prime \rangle |^2 - |
\langle \bar{N}_H^\prime \rangle |^2)
\right]\ ,
\end{eqnarray}
where $\xi_0$ and $\xi_\pi$ are the Fayet-Iliopoulos $D$-terms of the
U(1)$^\prime$ symmetry on the branes at $y=0$ and $y=\pi R$,
respectively, $v^{\prime 3/2}$ is the VEV of the $\phi$ component
associated with the U(1)$^\prime$ symmetry, and $g^\prime$ is the gauge
coupling constant. If $\xi_0 \neq - \xi_\pi$, we obtain $| \langle
N_H^\prime \rangle |^2 - |\langle \bar{N}_H^\prime \rangle |^2 \neq 0$,
so that $v \neq 0$ is assured by Eq.(\ref{d-flat}). The singlet mode is
eaten by the U(1)$^\prime$ gauge field in this case.

In the vacuum with $v \neq 0$, the wave functions for the matter fields
${\bf 10}_i$, ${\bf \overline{5}}_i$, and ${\bf 1}_i$ are deformed in a
different way 
because of the $- \sqrt{2} g {\bf 16}^c_i {\bf 45}_\phi {\bf 16}_i$ terms
in the superpotential (\ref{superpotential}).
The bulk mass terms after the SO(10) breaking is modified as 
$m_i \to m_i - \sqrt{2} g_X Q_X v^{3/2} $,
where $Q_X$ is the U(1)$_X$ charges of 
${\bf 10}: -1$, ${\bf \overline{5}}: 3$, and ${\bf 1}: -5$.
This is the direct connection between the origin of the flavor hierarchy
and the SO(10) breaking.  As we see in the next Section, this
modification gives a realistic fermion-mass and mixing hierarchy.
The violation of the unwanted SU(5) relation in the Yukawa matrices $f_d
= f_e^{\rm T}$ is obtained from the effect of ${\bf 16}_i {\bf 16}_j
{\bf 10}_H \langle {\bf 45}_H \rangle$ operators
\cite{Albright:1998vf,Babu:1998wi}.

We can also consider the 5-dimensional models where the gauge symmetry
is broken by the boundary conditions
\cite{Kawamura:2000ev}.
First, we discuss the 5-dimensional models on the space-time $M^4 \times
S^1/(Z_2 \times Z_2^\prime)$, where the $Z_2$ and $Z_2^\prime$ break
SO(10) down to SU(4) $\times$ SU(2) $\times$ SU(2) gauge symmetry and 4D
$N=2$ SUSY down to $N=1$, respectively.
In this scenario,
the SO(10) gauge symmetry is explicitly broken down to
SU(4) $\times$ SU(2) $\times$ SU(2) 
on the brane at $y=0$
\cite{Dermisek:2001hp,Albright:2002pt,Kim:2002im}. 
We do not have to worry
about the doublet-triplet splitting problem since we can introduce the
Higgs field as $H:({\bf 1},\ {\bf 2},\ {\bf 2})$ representation field
and thus there is no color-triplet Higgs from the beginning
\cite{Hebecker:2001wq}. 
The SO(10)
breaking Higgs fields ${\bf 16}_H$ and ${\bf \overline{16}}_H$ also need
not to be the full multiplets and we assume the existence of $\Phi_H:({\bf
\bar{4}},\ {\bf 1},\ {\bf 2})$ and $\overline{\Phi}_H:({\bf 4},\ {\bf
1},\ {\bf 2})$,
and their VEVs break the gauge symmetry down to the Standard Model gauge
group.
The matter fields are also decomposed to 
$\Psi_L^i:({\bf 4},\ {\bf 2},\ {\bf 1})$ and 
$\Psi_R^i:({\bf \bar{4}},\ {\bf 1},\ {\bf 2})$ 
which are
required to originate from the different ${\bf 16}$ dimensional fields in
the bulk due to the $Z_2$ projection. 
By SU(4) $\times$ SU(2) $\times$ SU(2) invariant but SO(10)
non-invariant interactions, 
we obtain the violation of the unrealistic SU(5)
relation of $f_d = f_e^{\rm T}$ between the Yukawa matrix of the
down-type quarks $f_d$ and that of the charged leptons $f_e$.
For example, the following SO(10) non-invariant interaction terms do not
give a contribution to $f_d$ but to $f_e$ after the Standard Model
singlet components of $\Phi_H$ and $\overline{\Phi}_H$ acquire VEVs:
\begin{eqnarray}
 W \ni \frac{\delta (y)}{M}
\frac{a_{ij}}{M^2}\ 
(\Psi_L^i  \Phi_H)_{({\bf 1},{\bf 2},{\bf 2})}
(\Psi_R^j \overline{\Phi}_H)_{({\bf 1},{\bf 1},{\bf 3})} 
H_{({\bf 1},{\bf 2}, {\bf 2})} \ ,
\end{eqnarray}
where $a_{ij}$ are the dimensionless coupling constants.
Including these contributions, we do not have any relation between the
down-type-quark and lepton Yukawa matrices.
Since the contributions should be comparable to the SO(10) invariant
terms of $\Psi_L \Psi_R H$ in order to reproduce the fermion masses, the
parameter $M$ is required to be the same scale as the VEV of
$\Phi_H$. 
Therefore, we take all the dimensionfull parameters to be of
the order of the GUT scale $\sim 10^{16}$ GeV.

In addition, we can discuss the scenario where the $Z_2$ and
$Z_2^\prime$ break 4D $N=2$ SUSY down to $N=1$ and SO(10) down to SU(5)
$\times$ U(1)$_X$, respectively.
Because of the $Z_2^\prime$ projection, we introduce six copies of ${\bf
16}$ fields in the bulk.
In this model, ${\bf 10}_i$ and (${\bf \bar{5}}_i$ and ${\bf 1}_i$) come
from the different ${\bf 16}$'s, and then, can have different bulk masses.
So, similar to SU(5) models in Ref.\
\cite{Hebecker:2002re},
one can generate the fermion-mass hierarchy by choosing a parameter set
$m_i$ ($i=1-6$) for six ${\bf 16}$ representation matter fields.
The doublet-triplet splitting problem can be solved by
Dimopoulos-Wilczek mechanism.
Essentially speaking, this is an SU(5) model but not SO(10) model,
and then the matter unification is spoiled. Thus, we will not discuss
it in this paper.

\section{Fermion masses}

We consider the fermion masses in this scenario with the assumption that
all the components of the Yukawa matrices on the brane are of order
unity. We also assume that neither SO(10) nor SU(5) relations among the
Yukawa matrices maintain because of the symmetry breaking effects on
the brane as discussed in Subsection 2.3.
The hierarchical structures in the observed fermion masses and mixings
are realized by means of the wave-function profiles of the zero modes
discussed in Subsection 2.1.
The lopsided family structure is given through the effect of
non-vanishing $v^{3/2} = \langle {\bf 45}_\phi \rangle$ in the bulk.

The observed hierarchy can be approximately expressed by the power of
the Cabibbo angle $\lambda \sim 0.22$.
The CKM matrix is given by
\begin{eqnarray}
 V_{\rm CKM} \sim \left(
\begin{array}{ccc}
 1 & \lambda & \lambda^3 \\
 -\lambda& 1 & \lambda^2 \\
 -\lambda^3& -\lambda^2 & 1\\
\end{array}
\right)\ .
\label{ckm}
\end{eqnarray}
The ratios of the masses at the GUT scale 
\cite{Fusaoka:1998vc}
are given by
\begin{eqnarray}
 m_u : m_c : m_t \sim \lambda^7 : \lambda^4 : 1 \ ,
\end{eqnarray}
\begin{eqnarray}
 m_d : m_s : m_b \sim \lambda^4 : \lambda^2 : 1 \ ,
\end{eqnarray}
\begin{eqnarray}
 m_e : m_\mu : m_\tau \sim \lambda^5 : \lambda^2 : 1 \ .
\end{eqnarray}
Also, the ratio of the neutrino masses in the
second and third generations is
estimated from the ratio of the $\Delta m^2$ of the solar and
atmospheric neutrino oscillations. If we assume the hierarchical mass
pattern, the ratio is given by
\begin{eqnarray}
 m_{\nu_2} : m_{\nu_3} \sim 
\sqrt{\frac{\Delta m^2_{\rm solar}}{\Delta m^2_{\rm atm.}}}
\sim \lambda^{1-2} : 1\ ,
\label{neutrino-ratio}
\end{eqnarray}
where we used the LMA solution for the solar neutrino oscillation which
requires nearly bimaximal mixing in the Maki-Nakagawa-Sakata (MNS)
matrix 
\cite{Maki:1962mu}
as follows:
\begin{eqnarray}
 V_{\rm MNS} \sim \left(
\begin{array}{ccc}
 1/\sqrt{2}& -1/\sqrt{2} & \epsilon \\
 1/2 & 1/2 & -1/\sqrt{2} \\
 1/2 & 1/2 & 1/\sqrt{2} \\
\end{array}
\right)\ ,
\label{mns}
\end{eqnarray}
where $\epsilon$ is a small parameter.

The Yukawa couplings in the effective theory are the products of the
Yukawa couplings and the values of the wave functions on the brane at
$y=0$ ($f_0(0)$). 
From Eq.(\ref{zero-mode}), the $f_0 (0)$'s are
given by
\begin{eqnarray}
 f_0(0)^{i,(r)} = \sqrt{\frac{2 |M| a_i^{(r)}}{1 - e^{-2 a_i^{(r)} c}}}
\sim \left \{ 
\begin{array}{ll}
 \sqrt{ 2|M a_i^{(r)}|} \ e^{-|a_i^{(r)}| c} & (a_i^{(r)} < 0)\\
 \sqrt{2|M| a_i^{(r)}} & (a_i^{(r)} > 0)\\
 \sqrt{|M|/c} & (a_i^{(r)} = 0) \\
\end{array}
\right.\ ,
\label{N-factor}
\end{eqnarray}
where the superscript $r$ stands for the belonging SU(5) representations
such as ${\bf 10}$, ${\bf \overline{5}}$, and ${\bf 1}$, and $i$ is the
generation index. The parameter $c$ is $|M| \pi R$. The $a_i^{(r)}$
parameters are dimensionless quantities expressed in terms of $\mu_i
\equiv m_i / |M|$ and $k \equiv g_X v^{3/2} / |M|$ as follows:\footnote{
In the case where SO(10) is broken 
down to SU(4) $\times$ SU(2) $\times$ SU(2)
by the boundary
condition as discussed in the previous section, the left-handed
multiplet $\Psi_L^i$ and right-handed one $\Psi_R^i$ may have
different bulk masses $m_i^L$ and $m_i^R$. However, as we see later,
only the case with $m_i^L \sim m_i^R$ gives the correct mass patterns,
and this is suitable to the horizontal unification and the gauge
coupling unification.  }
\begin{eqnarray}
 a_i^{({\bf 10})} = \mu_i + k,\ \ \ 
 a_i^{({\bf \overline{5}})} = \mu_i - 3 k,\ \ \ 
 a_i^{({\bf 1})} = \mu_i + 5 k\ .
\label{a-parameter}
\end{eqnarray} 
The parameter $k$ represents the effect of SO(10) breaking and the
SO(10) relation of $a_i^{({\bf 10})} = a_i^{({\bf \overline{5}})} =
a_i^{({\bf 1})}$ restores in the limit of $k \to 0$.  This parameter
plays an important role to reproduce the fermion-mass hierarchy.
Hereafter, we take $|\mu_3| = 1$ which can be done without loss of
generality.
Defining the dimensionless suppression factors $n_i^{(R)}
\equiv f_0(0)^{i,(r)} / \sqrt{|M|}$, we obtain the formulae of the Yukawa
matrices:
\begin{eqnarray}
 f_u^{ij} = \lambda_u^{ij} n_i^{({\bf 10})} n_j^{({\bf 10})},\ \ \ 
 f_d^{ij} = \lambda_d^{ij} n_i^{({\bf 10})} n_j^{({\bf \overline{5}})},
\ \ \ 
 f_e^{ij} = \lambda_e^{ij} n_i^{({\bf \overline{5}})} n_j^{({\bf 10})}\ .
\label{yukawa}
\end{eqnarray}
The matrices $\lambda_u$, $\lambda_d$, and $\lambda_e$ are the Yukawa
matrices on the brane at $y=0$
and their components are of order unity.
For neutrinos, the Yukawa matrix and the Majorana mass matrix for the
right-handed neutrinos are given by
\begin{eqnarray}
 f_\nu^{ij} = \lambda_\nu^{ij} n_i^{({\bf \overline{5}})} n_j^{({\bf 1})},
\ \ \ 
M_R^{ij} = \widetilde{M}_R^{ij} n_i^{({\bf 1})} n_j^{({\bf 1})}\ .
\label{neutrino}
\end{eqnarray}
Again $\lambda_\nu$ and $\widetilde{M}_R$ are the Yukawa matrix and the
Majorana mass matrix ($\widetilde{M}_R = \tilde{\lambda} \langle {\bf
\overline{16}}_H \rangle^2 / M$) on the brane at $y=0$, respectively. 
Since the neutrino masses are given through 
the seesaw mechanism 
$m_\nu = f_\nu M_R^{-1} f_\nu^{\rm T}$
\cite{seesaw}, 
the suppression factors of $n_i^{({\bf 1})}$
cancel in this formula and thus the neutrino mass matrix is expressed by
\begin{eqnarray}
 m_\nu^{ij} = (\lambda_\nu \widetilde{M}_R^{-1} \lambda_\nu^{\rm T} )^{ij}
\ n_i^{({\bf \overline{5}})} n_j^{({\bf \overline{5}})} v_2^2\ ,
\label{neutrino-mass}
\end{eqnarray}
where $v_2 \sim 174$ GeV is the VEV of the Higgs field which couples to
the up-type quarks and neutrinos.

The formulae in Eqs.(\ref{yukawa}) and (\ref{neutrino}) are the same as
those of the FN mechanism except
the relations in Eq.(\ref{a-parameter}). The number of the free
parameters is reduced to be four which is $\mu_1$, $\mu_2$, $c$, and $k$
compared to the case of FN mechanism in which six charges (three $Q({\bf
10}_i)$ and three $Q({\bf \overline{5}}_i)$) and a parameter $\epsilon =
\langle \Phi \rangle /M$ are adjustable. Interestingly, we see in the
following that the four parameters are sufficient to reproduce the
Yukawa structures.

Our goal is to reproduce the lopsided family structures given as
follows:
\begin{eqnarray}
 f_u \sim \left(
\begin{array}{ccc}
 \lambda^6 & \lambda^5 & \lambda^3 \\
 \lambda^5 & \lambda^4 & \lambda^2 \\
 \lambda^3 & \lambda^2 & 1\\
\end{array}
\right),\ \ \ 
 f_d \sim f_e^{\rm T} \sim \left(
\begin{array}{ccc}
 \lambda^4 & \lambda^3 & \lambda^3 \\
 \lambda^3 & \lambda^2 & \lambda^2 \\
 \lambda & 1 & 1\\
\end{array}
\right),\ \ \ 
 m_\nu \propto \left(
\begin{array}{ccc}
 \lambda^2 & \lambda & \lambda \\
 \lambda & 1 & 1 \\
 \lambda & 1 & 1\\
\end{array}
\right)\ .\ \ 
\end{eqnarray}
We can see that this form of Yukawa matrices reproduce the hierarchical
structures in Eqs.(\ref{ckm})-(\ref{mns}) when we take into account the
$O(1)$ ambiguities.
The large mixing in the 1-2 generation of the neutrinos is realized when
the determinant of the 2-3 submatrix in $m_\nu$ is $O(\lambda)$, which
is also consistent with the mass relation in Eq.(\ref{neutrino-ratio}).
As mentioned in Introduction, the form is suitable to the FN
mechanism.  Similarly, in our mechanism, the above structures are given
when the suppression factors have the form of
\begin{eqnarray}
 n_i^{({\bf 10})} \sim (\lambda^3, \lambda^2, 1), \ \ \ 
 n_i^{({\bf \overline{5}})} \sim (\lambda, 1, 1)\ .
\end{eqnarray}

\begin{figure}[t]
\includegraphics[width=15cm]{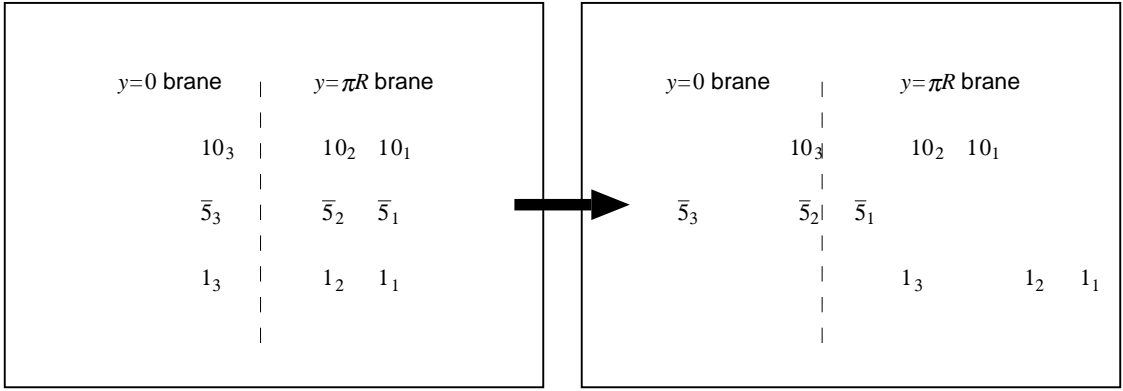} 
\caption{The configurations of the matter fields before and after the SO(10)
 breaking are shown.
\label{fig-1}}
\end{figure}

The above type of suppression factors is obtained in the following
way. First, we consider the $n_i^{({\bf 10})}$ factors. The last
component in $n_i^{({\bf 10})}$
of $O(1)$ is given if ${\bf 10}_3$ is localized on the
$y=0$ brane. From Eq.(\ref{N-factor}), this is achieved by 
\begin{eqnarray}
 a_3^{({\bf 10})} = \pm 1 + k \geq 0\ ,
\end{eqnarray}
where $\pm$ represent the undetermined sign of $\mu_3$.
The second component in $n_i^{({\bf 10})}$
of $O(\lambda^2)$ is, in contrast, given by the ${\bf 10}_2$
localization on the $y=\pi R$ brane such that
\begin{eqnarray}
 a_2^{({\bf 10})} c = (\mu_2 + k ) c \sim \log \lambda^2 \sim -3.0 \ .
\label{mu2}
\end{eqnarray}
The first component in $n_i^{({\bf 10})}$
of $O(\lambda^3)$ determine the relative size of the
$\mu_1$ parameter as follows:
\begin{eqnarray}
 (a_1^{({\bf 10})} - a_2^{({\bf 10})} ) c = (\mu_1 - \mu_2) c
\sim \log
 \frac{O(\lambda^3)}{O(\lambda^2)} \sim -1.5\ .
\label{mu1}
\end{eqnarray}
Next, the $n_i^{({\bf \overline{5}})}$ factors constrain the
parameters. The second and last components in 
$n_i^{({\bf \overline{5}})}$ of $O(1)$ give
\begin{eqnarray}
 a_3^{({\bf \overline{5}})} = \pm 1 - 3k \geq 0\ ,\ \ \ 
 a_2^{({\bf \overline{5}})} = \mu_2 - 3k \geq 0\ .
\end{eqnarray}
The suppression of the first component in $n_i^{({\bf \overline{5}})}$
is given by
\begin{eqnarray}
 a_1^{({\bf \overline{5}})} = (\mu_1 - 3k) c \sim \log \lambda \sim -1.5 \ .
\label{a2-5bar}
\end{eqnarray}
From the above conditions, 
the sign of $\mu_3$ is determined to be $\mu_3 = +1$ 
and the $k c$ parameter is approximately given by
\begin{eqnarray}
 k c \sim -0.75\ .
\label{scenario1}
\end{eqnarray}
Once we postulate the value of $k$ or $c$, we can determine the values
of $\mu_1$ and $\mu_2$ by Eqs.(\ref{mu2}) and (\ref{mu1}). For example,
for $k \sim -0.75$ and $c \sim 1$, we find $\mu_1 \sim -3.75$ and $\mu_2
\sim -2.25$.
We schematically explain this mechanism in Fig.{\ref{fig-1}}. 
Without SO(10) breaking, ${\bf 10}_i$ and ${\bf \overline{5}}_i$,
of course, have the same profiles. 
However, once the SO(10) breaking effect is
turned on, the profiles are deformed in such a way that the Yukawa
structures become the currently observed form.

When we think about the origin of the bulk mass parameters $m_i$, there
is a possibility that the origin is the spontaneous breaking of an SU(2)
horizontal symmetry\footnote{Similarly, one can consider the U(1)
flavor symmetry.}.
In fact, the suppression of the $n_1^{({\bf \overline{5}})}$ is not
necessarily important. Even if $n_1^{({\bf \overline{5}})} = O(1)$, the
Yukawa matrices are in a good shape if we take into account the $O(1)$
ambiguities in the Yukawa couplings on the brane. 
The neutrino masses
and mixings are explained by the anarchy scenario
\cite{Hall:1999sn}. 
In that scenario, the
form of the suppression factors is given by 
\cite{Hisano:2000wy}
\begin{eqnarray}
 n_i^{({\bf 10})} \sim (\lambda^4, \lambda^2, 1), \ \ \ 
 n_i^{({\bf \overline{5}})} \sim (1, 1, 1)\ .
\end{eqnarray}
Although this type of suppression factors gives a small value $\sim
\lambda^2$ to the Cabibbo angle, we think that the observed large value
of the Cabibbo angle is due to the $O(1)$ ambiguity.
An interesting possibility arises in this case. We find a parameter set
of $k \sim -1$, $c\sim 3$, $\mu_1 \sim -1$, $\mu_2 \sim 0$, and $\mu_3
=1$, which gives above suppression factors.
With the relation $\mu_1 = - \mu_3$ and $\mu_2=0$, there is a
possibility that the origin of the bulk mass parameters $m_i$ are the
VEV of the bulk adjoint field of the SU(2) horizontal symmetry in which the
three generations of the ${\bf 16}$ representation matter fields are
unified to ${\bf 3}$ representation.
As discussed in the previous section, the VEV of the bulk SU(2) adjoint
field gives rise to the $D$-terms on the branes and they are canceled by
the $D$-terms given by the VEVs of the localized fields 
in the fundamental representation on the branes. 
The VEVs break the SU(2) symmetry
completely and the couplings to the matter fields may induce off diagonal
components of the Yukawa couplings which are absent in the SU(2)
symmetric limit.
This possibility is a special feature in SO(10) GUT and can not be
realized in SU(5) GUT. The existence of the U(1)$_X$ breaking is
essential to give the different bulk masses to ${\bf 10}$ and ${\bf
\overline{5}}$ fields.

Finally, we comment on the neutrino masses. If we assume that all the
dimensionfull parameters are of the order of $10^{16}$ GeV, we find the
mass of the heaviest neutrino in Eq.(\ref{neutrino-mass}) is one order
of magnitude smaller than the favored value by the atmospheric neutrino
data of $\sim0.05$eV. However, this is not a serious problem because the
$O(1)$ ambiguities easily reduce the Majorana masses of the right-handed
neutrinos and/or increase the Dirac neutrino Yukawa couplings.

Although the right-handed neutrino masses do not affect the neutrino
masses, they might be important from the viewpoint of leptogenesis
\cite{Fukugita:1986hr}.
In the scenario with Eq.(\ref{scenario1}), our predictions on the
lightest right-handed neutrino masses are given by
\begin{eqnarray}
 M_R^{1} \sim M_R^{11} \sim 10^{-7} M \sim 10^{9} {\rm GeV}\ . 
\end{eqnarray}
This value is compatible with the leptogenesis scenario
\cite{Buchmuller:1998zf, Buchmuller:1996pa}.
Of particular interest is the case with the horizontal symmetry. With
the values of $k\sim -1$ and $c \sim 3$, we obtain the right-handed
neutrino masses $M_R^1 \sim 1-10$ GeV, $M_R^2 \sim 100-1000$ GeV, and
$M_R^3 \sim 10^{5-6}$ GeV.
Since the lightest right-handed neutrino does not enter the thermal
equilibrium before the electroweak phase transition, the
partial lepton asymmetry carried by the particles except for the
right-handed neutrino is converted to the baryon number by the sphaleron
process even if there is no net $B-L$ asymmetry 
\cite{Dick:1999je}.

We assumed in the superpotential in Eq.(\ref{superpotential}) that the
right-handed neutrinos acquire masses from the VEVs of $y=0$ brane fields.
Alternatively, we can change the source of their masses to the $y=\pi R$
brane.
However, in this case, 
Eq.(\ref{neutrino-mass}) is not applied,
and then it leads too small neutrino masses 
or non-suitable forms for the LMA solution.

\section{Discussions and Conclusions}

We consider the scenarios with the lopsided family structure in SO(10)
GUT by using the 5-dimensional wave function profile.
We can reproduce the fermion masses and mixings including the LMA
solution without introducing extra matter fields. The key point is that
we directly connect the SO(10) breaking effect to the origin of
fermion-mass hierarchy, which is difficult in the usual FN scenario.

We used the bulk mass terms as a origin of the hierarchy.  The SO(10)
breaking effect gives additional contributions to the bulk mass terms so
that the Yukawa matrices do not obey the unrealistic SO(10) relation
$f_u \sim f_d \sim f_e \sim f_\nu$. Moreover, we find the possibility of
SU(2) horizontal unification of three generations, and that is
possible in SO(10) rather than SU(5) GUT.

``SU(5) or SO(10)?'' is an interesting question. 
From the viewpoint of the doublet-triplet splitting problem,
SO(10) might have an advantage since it has an intrinsic
possibility to realize the doublet-triplet splitting by the
Dimopoulos-Wilczek mechanism, but it is not easy. 
Since we need to reduce the rank of gauge group, the Higgs sector
becomes complicated.
It is, therefore, difficult to judge from this viewpoint.
The fermion-mass hierarchy is another interesting viewpoint. We can say that
SU(5) is better than SO(10) in this case.  The SU(5) relation of $f_d
\sim f_e^{\rm T}$ is approximately good and the hierarchy is compatible
to the FN mechanism. On the other hand, SO(10) symmetry is
too strong and it is required to introduce extra-matter fields.
However, the concept of the matter unification is too beautiful to 
be thrown away.
In this sense, the wave-function localization is an attractive
possibility which enables a matter unified SO(10) GUT to be realistic
and opens up a further possibility of horizontal unification.

Considering the SUSY breaking in this scenario is also
interesting. Severe experimental limits of the processes with Flavor
Changing Neutral Current (FCNC) and CP violation restrict the sfermion
mass matrices to be some specific forms.
In our scenario, there is an interesting possibility if SUSY is broken
on the $y=\pi R$ brane. Because of the wave-function profiles, only the
first and second generation fields can feel the SUSY breaking effect
strongly. This is the desired situation for the effective SUSY scenario
\cite{Dine:1990jd} 
in which the sfermions of the first and second
generations have large soft masses of the order of 10 TeV so as to avoid
the constraints from the FCNC and CP violating processes, and the third
generation sfermions are as light as O(100) GeV for naturalness.
Consideration in this direction is interesting, although
we need the
mechanism to suppress the gaugino masses and to generate the $\mu$-term.
The gauge mediation 
\cite{Dine:1993yw}
is another candidate where 
the ${\bf 16}_H^\prime$ and ${\bf \overline{16}_H^\prime}$ fields 
on the $y=\pi R$ brane may be the messenger fields.

\section*{Acknowledgments}

We would like to thank Hooman Davoudiasl, Carlos Pena-Garay,
and Scott Thomas for useful discussions.
The work of R.K.\ was supported by DOE grant
DE-FG02-90ER40542.
The work of T.L.\ was supported by the National Science Foundation
under Grant No.~PHY-0070928.


\baselineskip 15pt
\begin{thebibliography}{99}
\bibitem{Witten:nf}
E.~Witten,
Nucl.\ Phys.\ B {\bf 188}, 513 (1981);\\
S.~Dimopoulos, S.~Raby and F.~Wilczek,
Phys.\ Rev.\ D {\bf 24}, 1681 (1981);\\
S.~Dimopoulos and H.~Georgi,
Nucl.\ Phys.\ B {\bf 193}, 150 (1981);\\
N.~Sakai,
Z.\ Phys.\ C {\bf 11}, 153 (1981).






\bibitem{Nilles:1983ge}
See, for example, H.~P.~Nilles,
Phys.\ Rept.\  {\bf 110}, 1 (1984).


\bibitem{Georgi:sy}
H.~Georgi and S.~L.~Glashow,
Phys.\ Rev.\ Lett.\  {\bf 32}, 438 (1974).



\bibitem{Giunti:ta}
C.~Giunti, C.~W.~Kim and U.~W.~Lee,
Mod.\ Phys.\ Lett.\ A {\bf 6}, 1745 (1991);\\
U.~Amaldi, W.~de Boer and H.~Furstenau,
Phys.\ Lett.\ B {\bf 260}, 447 (1991);\\
P.~Langacker and M.~x.~Luo,
Phys.\ Rev.\ D {\bf 44}, 817 (1991).



\bibitem{Albright:1998vf}
C.~H.~Albright, K.~S.~Babu and S.~M.~Barr,
Phys.\ Rev.\ Lett.\  {\bf 81}, 1167 (1998).

\bibitem{Babu:1998wi}
K.~S.~Babu, J.~C.~Pati and F.~Wilczek,
Nucl.\ Phys.\ B {\bf 566}, 33 (2000).


\bibitem{Wolfenstein:1977ue}
L.~Wolfenstein,
Phys.\ Rev.\ D {\bf 17}, 2369 (1978);\\
S.~P.~Mikheev and A.~Y.~Smirnov,
Sov.\ J.\ Nucl.\ Phys.\  {\bf 42}, 913 (1985)
[Yad.\ Fiz.\  {\bf 42}, 1441 (1985)];
Nuovo Cim.\ C {\bf 9}, 17 (1986).


\bibitem{Fukuda:2001nk}
S.~Fukuda {\it et al.}  [Super-Kamiokande Collaboration],
Phys.\ Rev.\ Lett.\  {\bf 86}, 5656 (2001);\\
Q.~R.~Ahmad {\it et al.}  [SNO Collaboration],
Phys.\ Rev.\ Lett.\  {\bf 89}, 011301 (2002);
Phys.\ Rev.\ Lett.\  {\bf 89}, 011302 (2002);\\
K.~Eguchi {\it et al.}  [KamLAND Collaboration],
Phys.\ Rev.\ Lett.\  {\bf 90}, 021802 (2003).



\bibitem{Fukuda:1998mi}
Y.~Fukuda {\it et al.}  [Super-Kamiokande Collaboration],
Phys.\ Rev.\ Lett.\  {\bf 81}, 1562 (1998).


\bibitem{Ahn:2002up}
M.~H.~Ahn {\it et al.}  [K2K Collaboration],
arXiv:hep-ex/0212007.



\bibitem{Froggatt:1978nt}
C.~D.~Froggatt and H.~B.~Nielsen,
Nucl.\ Phys.\ B {\bf 147}, 277 (1979).


\bibitem{Sato:1997hv}
J.~Sato and T.~Yanagida,
Phys.\ Lett.\ B {\bf 430}, 127 (1998);\\
T.~Yanagida and J.~Sato,
Nucl.\ Phys.\ Proc.\ Suppl.\  {\bf 77}, 293 (1999).



\bibitem{Buchmuller:1998zf}
W.~Buchmuller and T.~Yanagida,
Phys.\ Lett.\ B {\bf 445}, 399 (1999).

\bibitem{Irges:1998ax}
N.~Irges, S.~Lavignac and P.~Ramond,
Phys.\ Rev.\ D {\bf 58}, 035003 (1998).


\bibitem{Nomura:1998gm}
Y.~Nomura and T.~Yanagida,
Phys.\ Rev.\ D {\bf 59}, 017303 (1999);\\
Y.~Nomura and T.~Sugimoto,
Phys.\ Rev.\ D {\bf 61}, 093003 (2000);\\
C.~H.~Albright and S.~M.~Barr,
Phys.\ Rev.\ Lett.\  {\bf 85}, 244 (2000).

\bibitem{Maekawa:2001uk}
N.~Maekawa,
Prog.\ Theor.\ Phys.\  {\bf 106}, 401 (2001).

\bibitem{Kitano:2000xk}
R.~Kitano and Y.~Mimura,
Phys.\ Rev.\ D {\bf 63}, 016008 (2001).



\bibitem{Asaka:2001eh}
T.~Asaka, W.~Buchmuller and L.~Covi,
Phys.\ Lett.\ B {\bf 523}, 199 (2001);\\
L.~J.~Hall, Y.~Nomura, T.~Okui and D.~R.~Smith,
Phys.\ Rev.\ D {\bf 65}, 035008 (2002);\\
T.~Li,
Nucl.\ Phys.\ B {\bf 619}, 75 (2001);\\
B.~Kyae and Q.~Shafi,
arXiv:hep-ph/0212331.




\bibitem{Haba:2002ve}
N.~Haba and Y.~Shimizu,
arXiv:hep-ph/0210146;\\
N.~Haba, T.~Kondo and Y.~Shimizu,
Phys.\ Lett.\ B {\bf 535}, 271 (2002);
Phys.\ Lett.\ B {\bf 531}, 245 (2002).






\bibitem{Arkani-Hamed:1999dc}
N.~Arkani-Hamed and M.~Schmaltz,
Phys.\ Rev.\ D {\bf 61}, 033005 (2000).


\bibitem{Mirabelli:1999ks}
E.~A.~Mirabelli and M.~Schmaltz,
Phys.\ Rev.\ D {\bf 61}, 113011 (2000).



\bibitem{Kaplan:2001ga}
D.~E.~Kaplan and T.~M.~Tait,
JHEP {\bf 0111}, 051 (2001).


\bibitem{Kakizaki:2001ue}
M.~Kakizaki and M.~Yamaguchi,
arXiv:hep-ph/0110266.

\bibitem{Haba:2002uw}
N.~Haba and N.~Maru,
Phys.\ Rev.\ D {\bf 66}, 055005 (2002).


\bibitem{Arkani-Hamed:2001tb}
N.~Arkani-Hamed, T.~Gregoire and J.~Wacker,
JHEP {\bf 0203}, 055 (2002).


\bibitem{Hebecker:2002re}
A.~Hebecker and J.~March-Russell,
Phys.\ Lett.\ B {\bf 541}, 338 (2002).



\bibitem{Dimopoulos-Wilczek}
S.~Dimopoulos and F.~Wilczek, NSF-ITP-82-07.

\bibitem{Kawamura:2000ev}
Y.~Kawamura,
Prog.\ Theor.\ Phys.\  {\bf 105}, 999 (2001);\\
G.~Altarelli and F.~Feruglio,
Phys.\ Lett.\ B {\bf 511}, 257 (2001);\\
L.~J.~Hall and Y.~Nomura,
Phys.\ Rev.\ D {\bf 64}, 055003 (2001).




\bibitem{Hebecker:2001wq}
A.~Hebecker and J.~March-Russell,
Nucl.\ Phys.\ B {\bf 613}, 3 (2001).


\bibitem{Pomarol:1998sd}
A.~Pomarol and M.~Quiros,
Phys.\ Lett.\ B {\bf 438}, 255 (1998).




\bibitem{Barbieri:2002ic}
R.~Barbieri, R.~Contino, P.~Creminelli, R.~Rattazzi and C.~A.~Scrucca,
Phys.\ Rev.\ D {\bf 66}, 024025 (2002);\\
S.~Groot Nibbelink, H.~P.~Nilles and M.~Olechowski,
Phys.\ Lett.\ B {\bf 536}, 270 (2002);
Nucl.\ Phys.\ B {\bf 640}, 171 (2002);\\
H.~Abe, T.~Higaki and T.~Kobayashi,
arXiv:hep-th/0210025.




\bibitem{Barr:1997hq}
S.~M.~Barr and S.~Raby,
Phys.\ Rev.\ Lett.\  {\bf 79}, 4748 (1997).


\bibitem{Babu:1993we}
K.~S.~Babu and S.~M.~Barr,
Phys.\ Rev.\ D {\bf 48}, 5354 (1993);\\
J.~C.~Pati,
arXiv:hep-ph/0204240.



\bibitem{Dermisek:2001hp}
R.~Dermisek and A.~Mafi,
Phys.\ Rev.\ D {\bf 65}, 055002 (2002).

\bibitem{Albright:2002pt}
C.~H.~Albright and S.~M.~Barr,
Phys.\ Rev.\ D {\bf 67}, 013002 (2003).

\bibitem{Kim:2002im}
H.~D.~Kim and S.~Raby,
JHEP {\bf 0301}, 056 (2003).



\bibitem{Fusaoka:1998vc}
H.~Fusaoka and Y.~Koide,
Phys.\ Rev.\ D {\bf 57}, 3986 (1998).



\bibitem{Maki:1962mu}
Z.~Maki, M.~Nakagawa and S.~Sakata,
Prog.\ Theor.\ Phys.\  {\bf 28}, 870 (1962).



\bibitem{seesaw}
T.~Yanagida,
in
{\em Proceedings of the Workshop on Unified Theory
and Baryon Number of the Universe},
eds. O. Sawada and A. Sugamoto (KEK, 1979) p.95;\\
M.~Gell-Mann, P.~Ramond, and R.~Slansky,
in {\em Supergravity},
eds. P.~van Nieuwenhuizen and D.~Freedman
(North Holland, Amsterdam, 1979).


\bibitem{Hall:1999sn}
L.~J.~Hall, H.~Murayama and N.~Weiner,
Phys.\ Rev.\ Lett.\  {\bf 84}, 2572 (2000);\\
N.~Haba and H.~Murayama,
Phys.\ Rev.\ D {\bf 63}, 053010 (2001);\\
A.~de Gouvea and H.~Murayama,
arXiv:hep-ph/0301050.


\bibitem{Hisano:2000wy}
J.~Hisano, K.~Kurosawa and Y.~Nomura,
Nucl.\ Phys.\ B {\bf 584}, 3 (2000).




\bibitem{Fukugita:1986hr}
M.~Fukugita and T.~Yanagida,
Phys.\ Lett.\ B {\bf 174}, 45 (1986).

\bibitem{Buchmuller:1996pa}
W.~Buchmuller and M.~Plumacher,
Phys.\ Lett.\ B {\bf 389}, 73 (1996).


\bibitem{Dick:1999je}
K.~Dick, M.~Lindner, M.~Ratz and D.~Wright,
Phys.\ Rev.\ Lett.\  {\bf 84}, 4039 (2000);\\
H.~Murayama and A.~Pierce,
Phys.\ Rev.\ Lett.\  {\bf 89}, 271601 (2002).



\bibitem{Dine:1990jd}
M.~Dine, A.~Kagan and S.~Samuel,
Phys.\ Lett.\ B {\bf 243}, 250 (1990);\\
S.~Dimopoulos and G.~F.~Giudice,
Phys.\ Lett.\ B {\bf 357}, 573 (1995);\\
A.~Pomarol and D.~Tommasini,
Nucl.\ Phys.\ B {\bf 466}, 3 (1996);\\
A.~G.~Cohen, D.~B.~Kaplan and A.~E.~Nelson,
Phys.\ Lett.\ B {\bf 388}, 588 (1996).


\bibitem{Dine:1993yw}
M.~Dine and A.~E.~Nelson,
Phys.\ Rev.\ D {\bf 48}, 1277 (1993);\\
M.~Dine, A.~E.~Nelson and Y.~Shirman,
Phys.\ Rev.\ D {\bf 51}, 1362 (1995);\\
M.~Dine, A.~E.~Nelson, Y.~Nir and Y.~Shirman,
Phys.\ Rev.\ D {\bf 53}, 2658 (1996).



\end{thebibliography}
\end{document}